\documentclass{article}
\usepackage[english]{babel}
\usepackage[T2A]{fontenc}
\usepackage[cp1251]{inputenc}
\usepackage[dvips]{graphicx}
\usepackage{amsmath}
\usepackage{amssymb}
\usepackage{latexsym}
\begin{document}
\rm
\let\sc=\sf
\begin{flushright}
Journal-Ref: Astronomy Letters, 2010, Vol. 36, No. 6, pp. 422–429
\end{flushright}
\begin{center}
\LARGE {\bf Bimodal Brightness Oscillations in Models of Young Binary Systems}\\

\vspace{1cm}
\Large {\bf T. V.\,Demidova$^{1,2}$, N. Ya.\,Sotnikova$^1$
V. P.\,Grinin$^{1,2}$}

\normalsize \vspace{5mm}
1 - Sobolev Astronomical Institute, St. Petersburg State University, Universitetskii pr. 28, St. Petersburg, 198504 Russia,

2 - Pulkovo Astronomical Observatory, Russian Academy of Sciences, Pulkovskoe shosse 65, St. Petersburg,
                                         196140 Russia, \\
\end{center}
Received November 17, 2009
\normalsize
\begin{abstract}
We consider a model for the cyclic activity of young binary stars that accrete matter from the remnants of a protostellar cloud. If the orbit of such a binary system is inclined at a small angle to the line of sight, then the streams of matter and the density waves excited in the circumbinary disk can screen the primary component of the binary from the observer. To study these phenomena by the SPH (smoothed particle hydrodynamics) method, we have computed grids of hydrodynamic models for
binary systems based on which we have constructed the light curves as a function of the orbital phase.The main emphasis is on investigating the properties of the brightness oscillations. Therefore, the model parameters were varied within the following ranges: the component mass ratio $q = M_2 : M_1 = 0.2$ – $0.5$ and the eccentricity $e = 0$ – $0.7$. The parameter that defined the binary viscosity was also varied. We adopted optical grain characteristics typical of circumstellar dust. Our computations have shown that bimodal oscillations are excited in binaries with eccentric orbits, provided that the binary components do not differ too much in mass. In this case, the ratios of the periods and amplitudes of the bimodal oscillations and their shape depend strongly on the inclination of the binary plane and its orientation relative to the observer. Our analysis shows that the computed light curves can be used in interpreting the cyclic activity of UX Ori stars.

\end{abstract}
\clearpage

\clearpage
\large
\newpage
\section{INTRODUCTION}

 The circumstellar extinction variations due to an
inhomogeneous structure of the circumstellar gas–
dust disks are among the causes of the photometric
variability of young stars. This variability mechanism
dominates in UX Ori stars, because their circumstellar
disks have an “optimal” orientation at which the
line of sight “touches” the surface of the disk or, more
precisely, passes through its inhomogeneous gas–dust atmosphere (Grinin et al. 1991). For such a disk
orientation, the current brightness state of the star is
determined by the amount of dust on the line of sight
at the time of its observation. The amount of dust
on the line of sight changes continuously due to the
differential rotation of the matter in the disk, which
has an effect on the stellar brightness.

Analysis of long series of photometric observations
shows that apart from irregular circumstellar
extinction variations, many stars of this type also
exhibit wavelike (cyclic) variations with characteristic
time scales from several to ten or more years
(Shevchenko et al. 1993; Grinin et al. 1998; Rostopchina
et al. 1999; Herbst and Schevchenko 1999;
Bertout 2000; Shakhovskoi et al. 2005). In some
cases, there are simultaneously two cycles with different
periods in the light variations of UX Ori stars.

Cyclic photometric variability reflects the presence
of large-scale gas–dust structures in the circumstellar
disks. Grinin et al. (1998) and Rostopchina
et al. (1999) put forward the idea that the density
waves and the streams of matter attributable to the
presence of a secondary component or a protoplanet
are such structures. Their existence is predicted by
the hydrodynamic models of young binary systems
computed by Artymowicz and Lubow (1996). The
computations of these authors showed that a matter free
gap is formed in the central part of a young binary
system under gravitational perturbations. Its size depends
on the component mass ratio of the binary
and its orbital eccentricity. Under viscous forces and
gravitational perturbations, two streams of matter
generally unequal in intensity penetrate this gap; they
maintain the accretion activity of the components and
produce accretion disks around each of them.                                                        

The hydrodynamic computations by Sotnikova
and Grinin (2007) showed that three types of extinction
variations could arise on the line of sight
in young binary systems accreting matter from the
remnants of a protostellar cloud and inclined at a
small angle to the line of sight. The shortest period $P_1$
is equal to the orbital one and is produced by the
streams of matter that periodically penetrate from the
circumbinary disk into the binary and cross the line of
sight. The second period $P_2$, which is approximately a
factor of $5$–$8$ longer than the orbital one, is produced
by the motion of a one-armed density wave in the
circumbinary disk. The third period $P_3$ is attributable
to slow precession of the circumbinary disk and is
equal to several hundred orbital periods.

In this paper, we study in detail the properties
of the first two oscillation modes. For this purpose,
we computed a family of hydrodynamic models for
young binary systems by the SPH method. Based
on these models, we constructed the light curves and
investigated the influence of the model parameters of
young binary systems on the properties of the first two
periods. The results of our analysis are discussed in
the context of the cyclic activity of UX Ori stars.
\section{THE COMPUTATIONAL METHOD}
We computed the hydrodynamic flows and density
waves in a young binary system using the
SPH algorithm described in detail by Sotnikova and
Grinin (2007). The mass of the circumbinary (CB)
disk was assumed to be low compared to the total
mass of the stars in the binary system, which allowed
the disk self-gravity to be neglected.We also assumed
that the disk was isothermal and that the dust was
well mixed with the gas. It should be noted that using
the isothermal approximation in our case is justifiable,
because the density perturbations in a comparatively
narrow region of the CB disk, near its inner boundary,
make a major contribution to the extinction variations
on the line of sight. This allows the sizes of the domain
of integration in the SPH method to be limited by a
radius of $20\,a$, where $a$ is the semimajor axis of the
orbit at eccentricity $e = 0$.

For each model, we computed the column density
of test particles toward the primary component of the
binary as a function of the orbital phase. As a rule,
the computations were performed for several hundred
orbital periods. This time interval is shorter than the
characteristic CB-disk dissipation time attributable
mainly to the accretion of matter onto the binary
components by a factor of several (Sotnikova and
Grinin 2007).

For the models computed in this way, we passed
from the column density of test particles to that of
real grains.We began our quantitative analysis of the
computational results with the removal of the trend
in the column density variations attributable to the
decrease in the number of test particles in the binary
as a result of their accretion onto its components. The
trend was modeled by a fifth-degree polynomial. Our
computations showed that this ensured a satisfactory
removal of the trend for all of the models considered.
Next, we performed a Fourier analysis of the time
series freed from the trend with the calculation of
the likelihood function with a probability $p = 0.99$
(Vityazev 2001). In this way, we revealed the cycles
of the model time series for the column density. The
period $P_2$ was determined from the centroid of the
corresponding peak in the Fourier spectrum.

To construct the bimodal light curves, we chose
a fragment of the particle column density variations
with a duration of $2\,P_2$ in each model. To reduce the
influence of fluctuations, we used the method of a
three-point moving average. The optical depth of the
dust layer on the line of sight should be determined
to pass from the column density of test particles to
magnitudes, which requires passing from the test
particles to real grains. This was done as follows:
first, we determined the “weight” of a single test
particle. For this purpose, we specified the accretion
rate onto the binary components $\dot M_a$ as a parameter
of the problem and compared this quantity with the
accretion rate of test particles determined during the
computations (for more detail, see Sotnikova and
Grinin 2007). Below, in our calculations, $\dot M_a$ was
taken to be $10^{-10}\, M_{\odot}$, which is approximately
an order of magnitude lower than the typical accretion
rates onto UX Ori stars (Tambovtseva et al. 2001;
Muzerolle et al. 2004). Our calculations showed that
for the orbital inclinations for which these were performed,
such an accretion rate is sufficient to produce
a strong modulation of the brightness of the primary
component in the binary.

The mass of a single particle was calculated as
follows:

\begin{equation}
m_{d}=\frac {P\,\dot{M_a}}{N} \label{md}
\end{equation}

where $P$ is the orbital period, $N$ is the total number
of test particles accreting on both components in one
binary revolution, and $\dot{M_a}$ is the adopted accretion
rate.

The area of the column section $s$ is a technical
parameter of the problem. It was specified as $s = 2h \times 2h$, 
where $h = 0.1\,a$. Our calculations showed
that this value of $s$ is optimal for the solution of the
formulated problem (at lower values of $s$, the influence
of fluctuations is enhanced; at its higher values,
the time resolution of the calculated characteristics
deteriorates). Thus, having calculated the mass of
a single test particle, we obtain the matter column
density in units of $g/cm^{2}$. Now, the opacity $\kappa$ per
gram of matter should be taken as a parameter of the
problem to determine the optical depth of the dust
on the line of sight. This parameter depends on the
type and sizes of the dust grains and on the dust-to-gas
mass ratio. The latter is commonly assumed to
be the same as that, on average, in the interstellar
medium: $1:100$. The typical circumstellar dust extinctions
in the optical spectral region lie within the
range $\kappa = 100-300\,cm^{2}/g$ (see, e.g., Natta and
Whitney 2000). In our calculations, we adopted $\kappa = 250\,cm^{2}/g$, 
which corresponds to the maximum of the Johnson B passband. Multiplying the matter
column densities calculated in this way by $\kappa$, we will
obtain the optical depths $\tau$ for each instant of time.

It should be noted that the intensity of the radiation
from young stars generally consists of two parts: the
intensity of the stellar radiation $I_{\ast}$ (in our case, the
primary component of the binary) attenuated by a factor
of $e^{- \tau}$ and the intensity of the radiation scattered
by circumstellar dust $I_{sc}$:

\begin{equation}
I_{obs}=I_{\ast}e^{-\tau} \label{Iobs} + I_{sc}\,,
\end{equation}

The contribution of the scattered light to the observed
radiation from young stars typically does not
exceed a fewpercent (the young HH 30 stars whose
direct radiation is severely attenuated by absorption
in their circumstellar disks seen edge-on constitute
an exception). Therefore, below, when studying the
pattern of variability in the primary component of
the binary, we took the intensity of the scattered
radiation in (2) to be zero. The light variations of
the primary component are expressed in magnitudes:
$\Delta m = -2.5\cdot lg\,I_{obs}$ ($I_*$ is taken as unity).

\section{RESULTS}
We computed theoretical light curves for binary
systems by the method described above. The basic
model parameters are given in the table. In all models,
the mass of the primary component is $2\,M_\odot$, a value
typical of many UX Ori stars (Rostopchina 1999);
$M_2$ is the mass of the secondary component; $c$ is
the dimensionless speed of sound in the matter expressed
in units of the orbital velocity of the secondary
component at $e = 0$. The number of test particles is
$N = 60 000$ and the smoothing length is $h = 0.1a$.
The orbital period is taken to be five years. For a
circular orbit, its radius is $3.7 AU$. The last column of
the table gives the values of the period $P_2$ calculated
by the method described above and expressed in units
of the orbital period.

The model parameters were varied within the
following ranges: the component mass ratio $q = M_2/M_1 = 0.1 - 0.5$ 
and the eccentricity $e = 0 - 0.7$; the dimensionless speed of sound in the CB disk $c$
appearing in the expression for viscosity was taken
to be $0.02$ (“cold” disk), $0.05$ (“warm” disk), and
$0.08$ (“hot” disk). Our computations were performed
for several inclinations of the equatorial plane of the
binary system to the line of sight and four positions
of the apsidal line relative to the observer (with a $90^\circ$
step). The choice of inclinations was restricted by a
finite number of test particles and the necessity of
avoiding great statistical fluctuations in the number
of particles on the line of sight.

\begin{table}[p]
\label{t_models}
\begin{center}
\caption{Parameters of the binary models.}
\begin{tabular}{c|c|c|c|c}
\hline \hline
  Model & $e$ & $M_2$ & $c$ & $P_2/P_1$\\
\hline
  1  & 0.3  & 0.7 & 0.05 & 4.73$\pm$0.07 \\
  2  & 0.5  & 0.7 & 0.05 & 5.93$\pm$0.06 \\
  3  & 0.7  & 0.7 & 0.05 & 6.77$\pm$0.14 \\
  4  & 0.5  & 1.0 & 0.08 & 5.97$\pm$0.01 \\
  5  & 0.5  & 1.0 & 0.05 & 6.56$\pm$0.10 \\
  6  & 0.5  & 0.4 & 0.05 & 5.19$\pm$0.16 \\
  7  & 0.1  & 0.7 & 0.05 & --- \\
  8  & 0.5  & 1.0 & 0.02 & 5.43$\pm$0.11\\
  9  & 0.5  & 0.2 & 0.05 & --- \\
\hline
\end{tabular}
\end{center}
\end{table}

Figures 1–4 present the bimodal light curves of a
binary system for different positions of the apsidal line
relative to the observer and two inclinations of the disk
plane to the line of sight.We see that when the orbital
orientation changes relative to the observer, the light
curve can change significantly. First, the shape of the
light curves changes with position of the apsidal line
and with orbital inclination. Second, the relationship
between the fast and slow oscillation modes changes.

The dependence of the light curves on the orbital
inclination is quite understandable if we take into account
the fact that the streams of matter propagating
from the CB disk to the central part of the binary
system contribute to the extinction variations with
orbital phase, while the period $P_2$ is produced by the
motion of a one-armed density wave in the CB disk.
For this reason, the contribution of this wave to the
extinction variations is more sensitive to changes in
the inclination and decreases rapidly as it increases.

The strong influence of the position of the apsidal
line on the light curves is caused by a global asymmetry
of the CB disk (for its origin, see Artymowicz
and Lubow1996). The azimuthal dependence of the
geometrical thickness of the inner CB disk region
(Sotnikova and Grinin 2007) results from this asymmetry.
It is this factor that is responsible for the strong
dependence of the binary’s light curves on the position
of the apsidal line.

In models 1–3 and 7, we varied the orbital eccentricity.
In model 7 ($e = 0.1$), there is no slow oscillation
mode with the period $P_2$. As $e$ increases from $0.3$
to $0.7$, the second period appears. In this case, the
ratio $P_2/P_1$ increases with $e$: $P_2/P_1 = 4.73 \pm 0.07$ in
model 1, $5.93 \pm 0.06$ in model 2, and $6.77 \pm 0.14$ in
model 3. This dependence of the slow oscillation mode
on $e$ can be explained as follows: the radius of the
inner gap in the CB disk increases with orbital eccentricity
(Artymowicz and Lubow 1994) and since the
one-armed density wave propagates with a velocity
close to the Keplerian one, increasing $e$ reduces the
velocity of its propagation over the CB disk.

We used models 2, 5, 6, and 9 to analyze the
influence of the binary component mass ratio on the
period ratio. Our computations showed that the amplitude
of the slow oscillation mode became vanishingly
small as $q$ decreased; the period itself $P_2$ also
decreased: in model 9 with the mass ratio $q = 0.1$,
there is no second period; the period ratio is $5.19 \pm
0.16$ in model 6 ($q = 0.2$), $5.93 \pm 0.06$ in model 2
($q = 0.35$), and $6.56 \pm 0.1$ in model 5 ($q = 0.5$). This
dependence of the period $P_2$ on the mass of the secondary
component is caused by the same factor as
that for the dependence of $P_2$ on eccentricity $e$: the
size of the inner gap in the CB disk increases with
mass of the secondary component (Artymowicz and
Lubow,1994), causing the angular velocity of the
density wave in the CB disk to decrease.

In models 4, 5, and 8, we varied the parameter $c$,
the dimensionless speed of sound in the matter that
characterizes the disk viscosity. Our computations
showed that the period $P_2$ reached its maximum at
$c = 0.05$. The period ratio is $P_2/P_1 = 5.97 \pm
0.01$ in model 4 ($c = 0.08$), $6.56 \pm 0.10$ in model 5 ($c =
0.05$), and $5.43 \pm 0.11$ in model 8 ($c = 0.02$). In this
case, the slow oscillation mode gradually degrades
with decreasing viscosity.

The results listed above were obtained without
taking into account the effect of scattered radiation.
Nevertheless, these are also valid in real systems with
a noticeable contribution of scattered light. This can
be seen from Fig. 5, which shows two light curves.
These were computed at $I_{sc} = 0$ and $I_{sc}$ = 0.1\,$I_{\ast}$. We
see that the scattered radiation reduces the amplitude
of the brightness oscillations, but their bimodal structure
is retained.

To conclude this section, the following should be
noted: in counting the number of test particles on
the line of sight, we had to limit ourselves to a small
range of orbital inclinations relative to the line of sight
($\le 10^\circ$), since the influence of statistical fluctuations
\section{DISCUSSION AND CONCLUSIONS}
The results presented above show that the streams
of matter and the density waves produced in a young
binary system by periodic gravitational perturbations
are capable of causing appreciable (in amplitude) periodic
extinction variations, which can lead to cyclic
light variations in young stars. The amplitude and
shape of the light curves depend significantly on the
binary parameters (component mass ratio, orbital eccentricity,
and disk viscosity), the accretion rate, and
the binary orientation in space. The latter includes not
only the inclination of the orbital plane to the line of
sight but also its orientation (for a noncircular orbit)
relative to the observer.

Our computations showed that two different cycles
could be simultaneously present in the light variations
of the binary’s primary component in a fairly
wide range of model parameters: one of them corresponds
to the orbital period and the other corresponds
to the revolution period of a one-armed density wave
near the inner CB disk boundary. The period ratio
$P_2/P_1$ depends on the binary parameters. It follows
from the table that in five of the seven models in which
both oscillation modes are simultaneously present,
the period of the second mode (expressed in fractions
of the orbital period) is close to an integer (5, 6, or 7).
This suggests that the Lindblad resonance closest to
the inner CB disk boundary plays an important role in
forming the slow cycle.

The amplitude of the brightness oscillations with
the period $P_2$ decreases with decreasing mass of the
secondary component and becomes vanishingly small
at $q \leq 0.1 $. It also decreases as one passes from eccentric
orbits to circular ones. Viscosity also has a
significant effect on the amplitude of both oscillations.

In the Introduction, we pointed out that the cyclic
activity of UX Ori stars is related to the extinction
variations presumably caused by the presence
of massive perturbing bodies (protoplanets, brown
dwarfs, binary components) in their neighborhoods.
Observations show that the amplitude of the cyclic
component in the light variations of these stars ranges
from several tenths of amagnitude to two magnitudes
in the $V$ band, i.e., it is comparable to the theoretical
values at a binary inclination of about $10^\circ$ (Figs. 1–4).
In fact, the circumstellar disks of UX Ori stars are inclined
relative to the line of sight at slightly larger angles
(see Natta andWhitney 2000). However, the accretion
rate for these stars is $10^{-8} M_\odot/yr$ (Tambovtseva
et al. 2001; Muzerolle et al. 2004), i.e., it is
higher than that in the models computed above by
a factor of 100. This suggests that the photometric
cycles can also be observed at binary inclinations
exceeding $10^\circ$ (see the previous section).

As was pointed out in the Introduction, two component
cycles were detected in some of the
UX Ori stars. In particular, for SV Cep (Rostopchina
et al. 1999) and CQ Tau (Shakhovskoi et al. 2005),
the ratios of the large and small cycles turned out
to be close to 6.1 and 7.5, respectively. Two cycles
(Rostopchina et al. 1999; Bertout et al. 2000) with a
ratio close to 7.5 were also detected in the star UXOri
itself. It should be noted that these values are still
determined from observations not very accurately,
because the series of photometric observations are
still insufficiently long in a number of cases. Nevertheless,
the closeness of all three period ratios to the
theoretical period ratios $P_2 : P_1$ given in the table is
immediately apparent. This suggests that the cyclic
activity of at least some UX Ori stars can actually
result from their binarity.

It should be noted that another object with bimodal
brightness oscillations exists among the variable
stars whose brightness varies periodically. This
is the carbon star V Hya. A detailed study of its
variability by Knapp et al. (1999) showed that out of
its two observed periods ($P_1 = 530^d$ and $P_2 = 6000^d$),
the short period is attributable to stellar pulsations,
while the long period is related to the star’s binarity
and is attributable to the eclipses by dust rotating
in the orbit together with the secondary component.
Some of the Herbig Ае stars, which the family of
UX Ori stars mostly consists of, also exhibit pulsations
(see Stahler and Palla (2004) and references
therein). However, the period of these pulsations is
too short (of the order of one hour) and, for this
reason, they cannot be involved in producing the bimodal
brightness oscillations observed in these stars.
Therefore, if the hydrodynamic processes in young
binary systems considered above are responsible for
these oscillations, then the shorter of the two periods
corresponds to the orbital period.

This makes it possible to estimate the amplitude
of the stellar radial velocity variations caused by the
orbital motion of the companion. Thus, for example,
the shorter of the two periods for CQ Tau is
2.7 yr. The mass of this star was estimated by Rostopchina
(1999) to be about $1.5\,M_\odot$. If the above
period corresponds to the orbital period of a companion
with a mass equal, for example, to 1/5 of the
mass of CQ Tau, then the amplitude of the stellar
radial velocity variations in this case will be about 10
km/s. Although the absorption lines in the spectra
of UX Ori stars are broadened by rapid rotation
and are severely distorted by variable circumstellar
absorption lines (Grinin et al. 2001), an attempt can
be made to measure these stellar radial velocity variations.

The deficit of infrared radiation in the energy distribution
(in the near infrared) due to the presence
of a matter-free central gap in the disk could be
other evidence for the binarity of a star. However,
this observational test cannot be considered quite
reliable, because much of the near-infrared radiation
from the circumstellar disks originates near the dust
evaporation zone. According to present views (Natta
et al. 2001), the circumstellar disks of young stars are
puffed up near this zone, producing excess infrared
radiation compared to the standard disk model. Concurrently,
this puffing-up produces a shadow zone in
the disk (Dullemond et al. 2001), causing a reduction
in the infrared radiation at longer wavelengths. The
existence of a shadow zone is similar in its observational
manifestations to the existence of a matter free
zone. Therefore, it is very different to obtain direct
observational evidence for the binarity of UX Ori
stars by measuring their radial velocities or from the
infrared energy distribution. Interferometric observations
with large interferometer telescopes like ALMA
will probably play an important role in solving this
problem.
\section{ACKNOWLEDGMENTS}
This was was supported by the “Origin and Evolution
of Stars and Galaxies” Program of the Presidium
of the Russian Academy of Sciences and the
Program for Support of Leading Scientific Schools
(NSh-1318.2008.2 and NSh-3645.2010.2). We are
grateful to the anonymous referee for helpful remarks.
\clearpage
\begin{center}
{\Large\bf REFERENCES}
\end{center}
1. P. Artymowicz and S.H. Lubow, Astrophys. J. 421, 651 (1994).\\
2. P. Artymowicz and S.H. Lubow, Astrophys. J. 467, L77 (1996).\\
3. C. Bertout, Astron. Astrophys. 363, 984 (2000).\\
4. C. P. Dullemon, C. Dominik, and A.Natta, Astrophys. J. 560, 957 2001.\\
5. V. P. Grinin, N. N. Kiselev, N. Kh. Minikulov, et al., Astrophys. Space Sci. 186, 283 (1991).\\
6. V. P. Grinin, A. N. Rostopchina, and D. N. Shakhovskoi, Pis’ma Astron. Zh. 24, 925 (1998) [Astron. Lett. 24, 802 (1998)].\\
7. V. Grinin, O. Kozlova, A. Natta, et al., Astron. Astrophys. 379, 482 (2001).\\
8. W. Herbst and V. S. Schevchenko, Astron. J. 118, 1043 (1999).\\
9. G. R. Knapp, S. I. Dobrovol’sky, Z. Ivezic, et al., Astron. Astrophys. 251, 97 (1999).\\
10. J. Muzerolle, P. D’Alessio, N. Calvet, and L. Hartmann, Astrophys. J. 617, 406 (2004).\\
11. A. Natta and B.Whitney, Astron. Astrophys. 364, 633 (2000).\\
12. A. Natta, T. Prusti, R. Neri, et al., Astron. Astrophys. 371, 186 (2001).\\
13. A. N. Rostopchina, Astron. Zh. 76, 136 (1999) [Astron. Rep. 43, 113 (1999)].\\
14. A. N. Rostopchina, V. P. Grinin, and D. N. Shakhovskoi, Pis’ma Astron. Zh. 25, 291 (1999) [Astron. Lett. 25, 243 (1999)].\\
15. V. S. Schevchenko, K. N. Grankin, M. A. Ibragimov, et al., Astrophys. Space Sci. 202, 137 (1993).\\
16. D. N. Shakhovskoj , V. P. Grinin, and A. N. Rostopchina, Astrophysics 48, 135 (2005).\\
17. N. Ya. Sotnikova and V. P. Grinin, Pis’ma Astron. Zh. 33, 667 (2007) [Astron. Lett. 33, 594 (2007)].\\
18. S. Staler and F. Palla, The Formation of Stars (Wiley-VCH,Weinheim, 2004).\\
19. L. V. Tambovtseva, V. P. Grinin, B. Rodgers, and O. Kozlova, Astron. Zh. 78, 514 (2001) [Astron. Rep.45, 442 (2001)].\\
20. V. V. Vityazev, Spectral–Correlation Analysis of Uniform Time Series (SPb. Gos. Univ., St. Petersbourg,2001) [in Russian].\\

Translated by V. Astakhov
\newpage

\begin{figure}[!h]\begin{center}
  \makebox[0.6\textwidth]{ \includegraphics[scale=1.3]{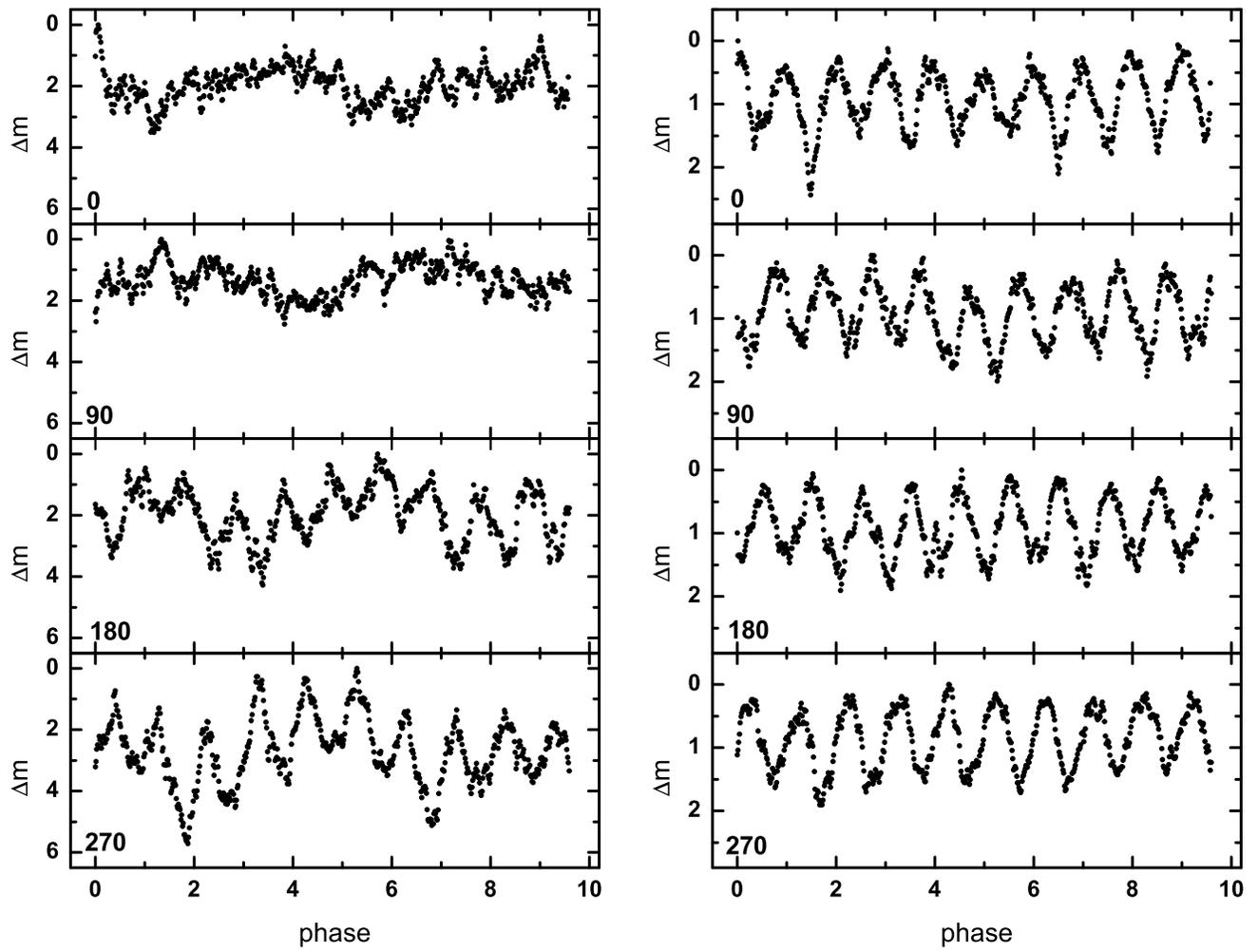}
  }
 \caption{ Bimodal brightness oscillations in model 1 (warm disk, e = 0.3, q = 0.35): (a) the line of sight lies in the orbital plane
and (b) the disk is inclined at an angle of $10^\circ$ to the line of sight. The angle that characterizes the position of the apsidal line
relative to the observer is indicated in the lower left corner of each panel.} \label{m_p1}
\end{center}
\end{figure}
\newpage
   \begin{figure}[!h]\begin{center}
  \makebox[0.6\textwidth]{ \includegraphics[scale=1.3]{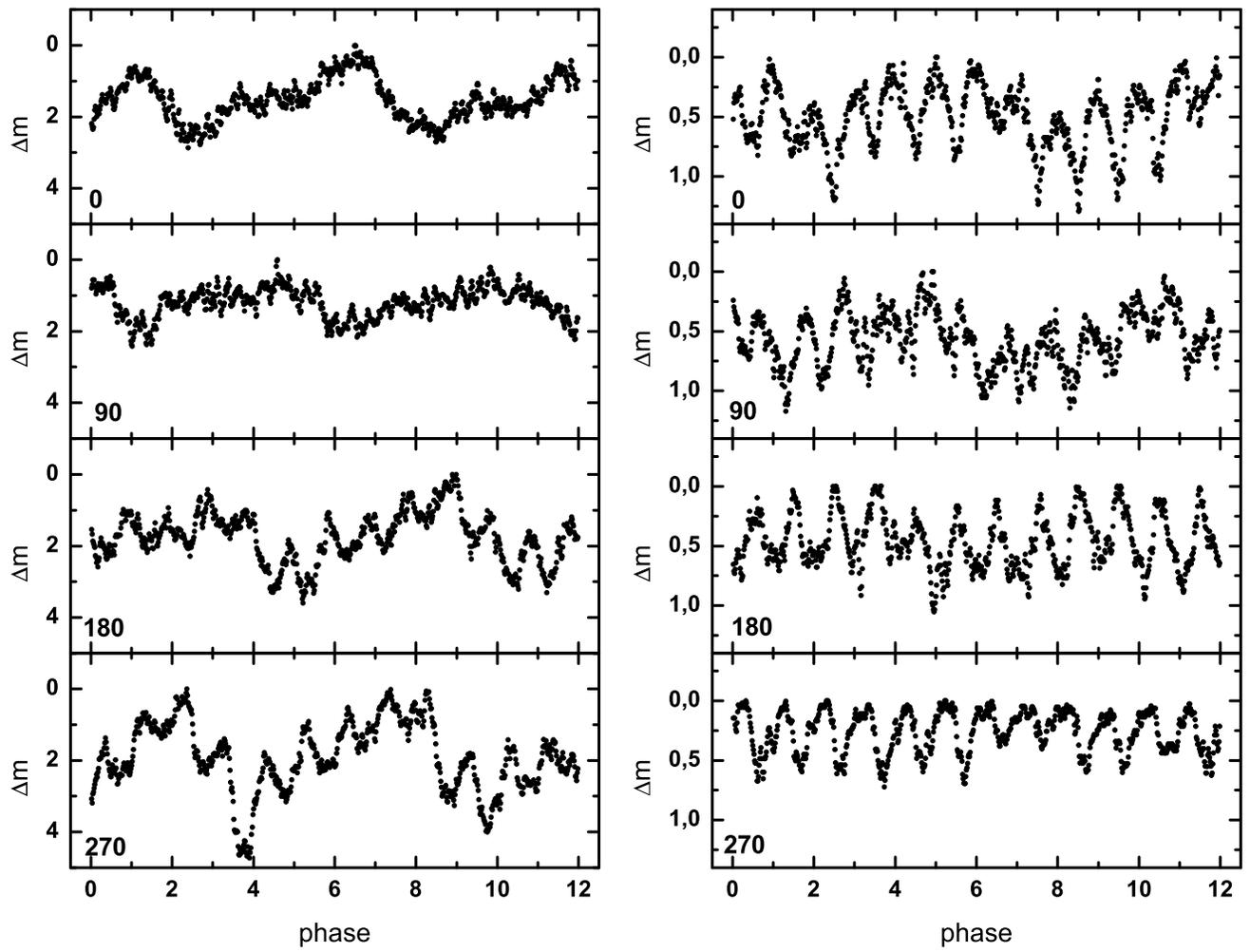}
  }
    \caption{Same as Fig. 1 for model 2 (warm disk, $e = 0.5$, $q = 0.35$).}
    \end{center}
   \end{figure}
\newpage
\begin{figure}[!h]\begin{center}
  \makebox[0.6\textwidth]{\includegraphics[scale=1.3]{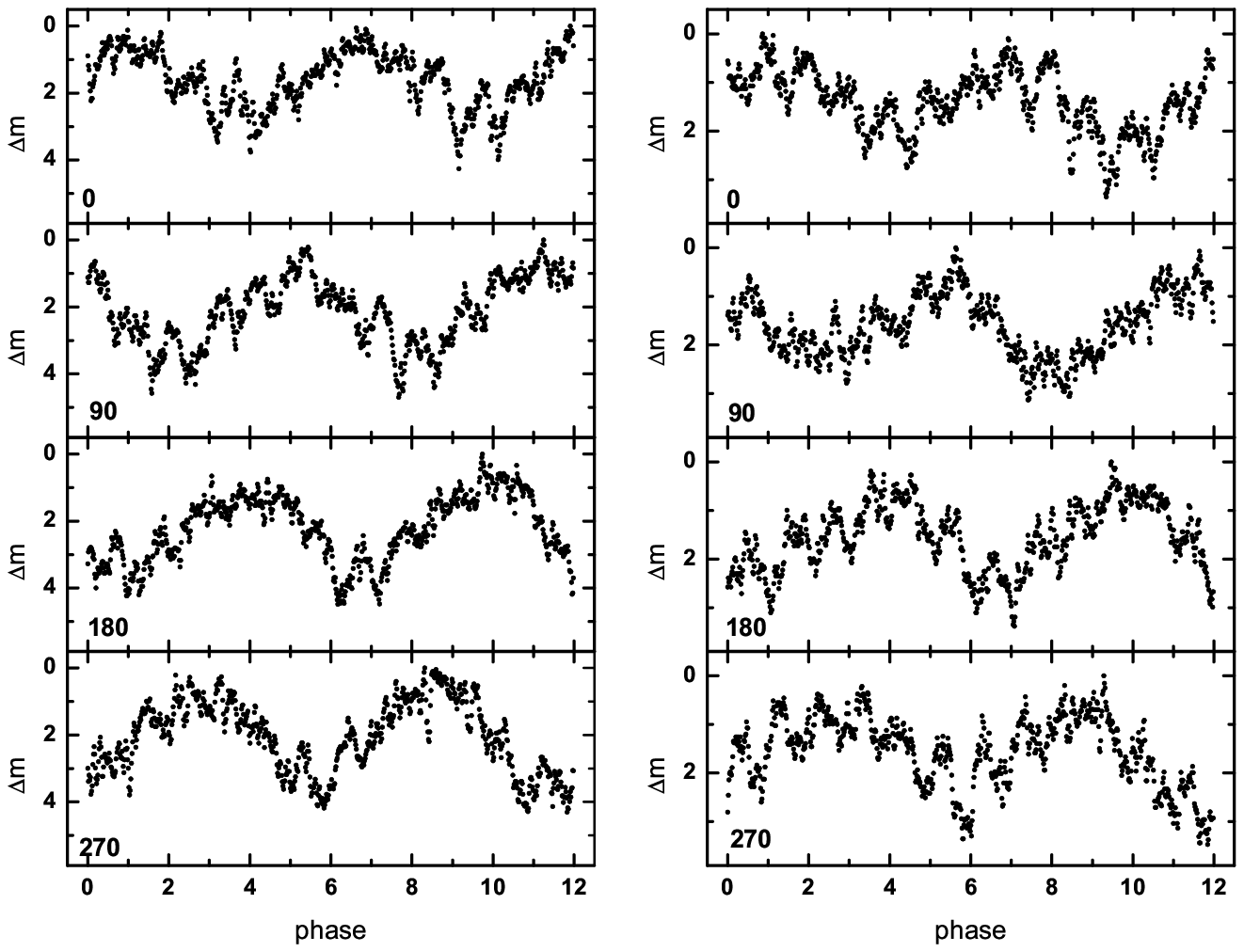}
  }
 \caption{Same as Fig. 1 for model 4 (hot disk, $e = 0.5$, $q = 0.5$). The light curves in the right part of the panel were constructed
for the case where the orbit was inclined at an angle of $7^\circ.5$ to the line of sight.
}
\end{center}
\end{figure}
\newpage
\begin{figure}[!h]\begin{center}
  \makebox[0.6\textwidth]{\includegraphics[scale=1.3]{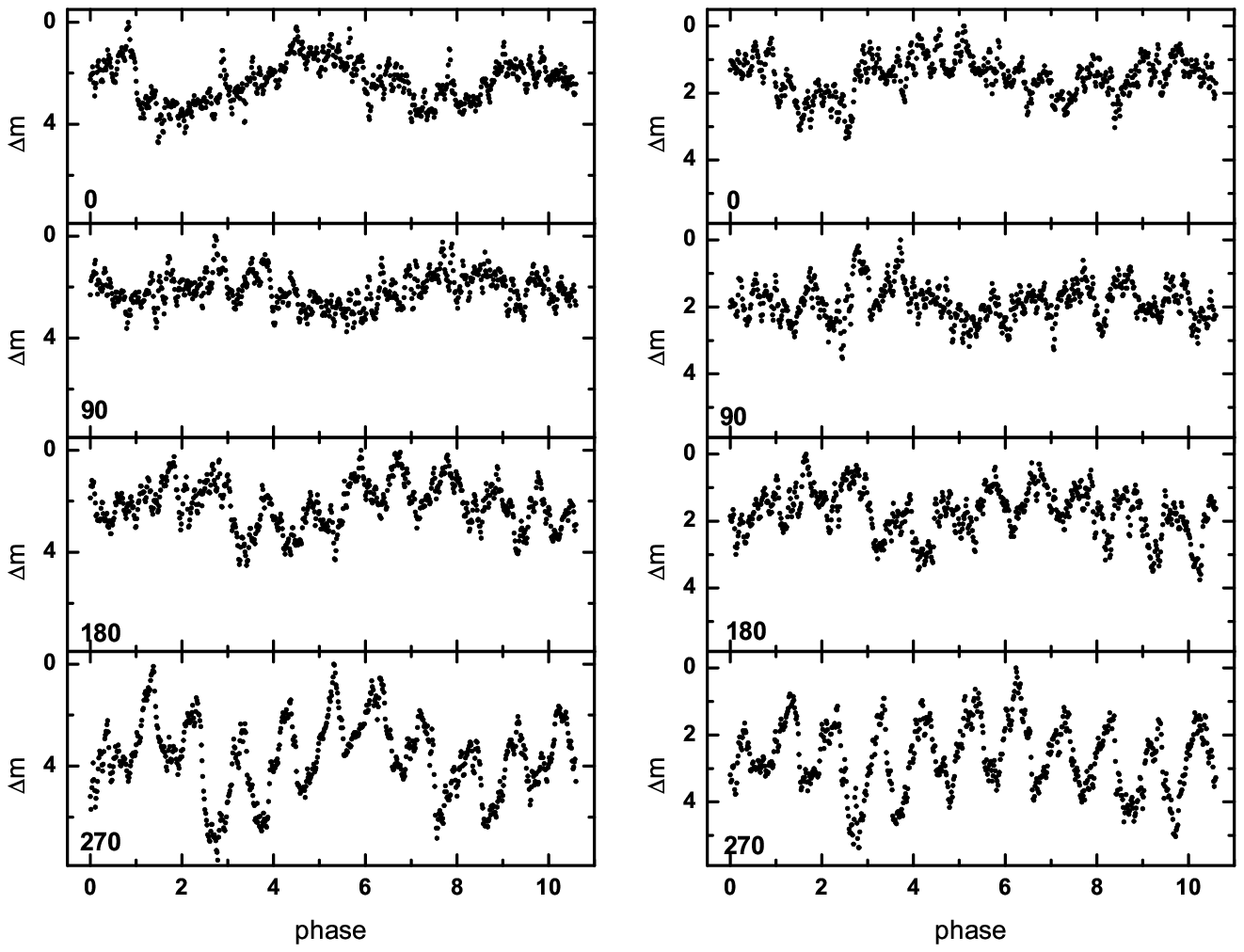}
  }
 \caption{Same as Fig. 1 for model 6 (warm disk, $e = 0.5$, $q = 0.2$). The light curves in the right part of the panel were constructed
for the case where the orbit was inclined at an angle of $5^\circ$ to the line of sight.}
\end{center}
\end{figure}
\newpage
\begin{figure}[!h]\begin{center}
  \makebox[0.6\textwidth]{\includegraphics[scale=1.3]{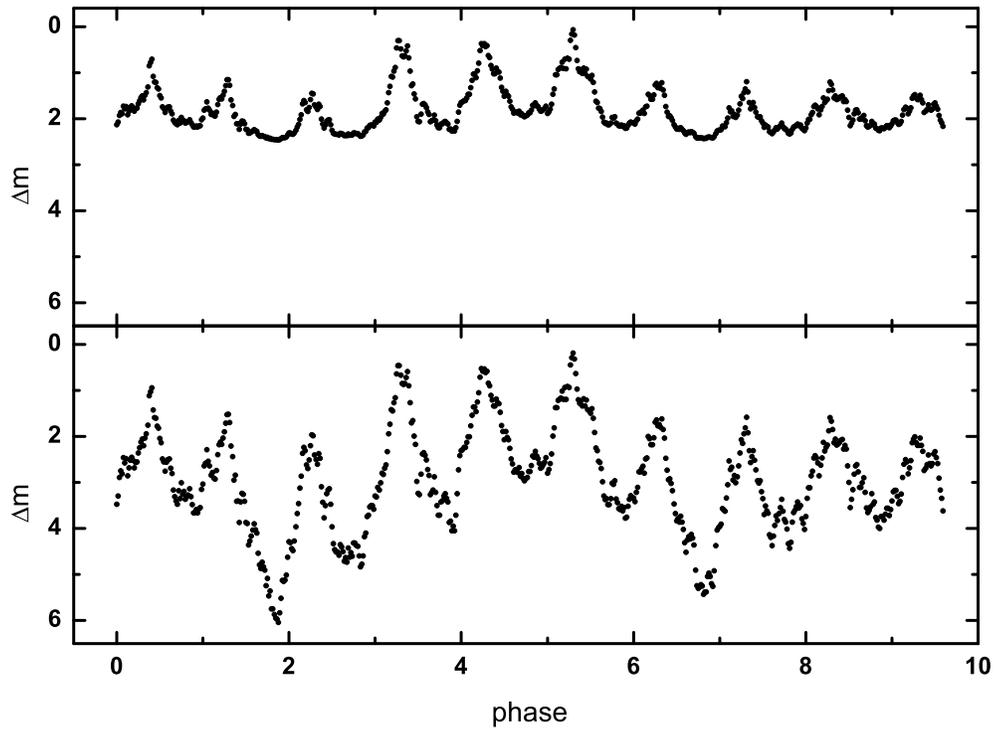}
  }
 \caption{ Light curves for model 1: the line of sight lies in the orbital plane and makes an angle of $270^\circ$ with the apsidal
line. The plot illustrates the influence of scattered light on the brightness modulation of the binary system: (a) $I_{sc} = 0.1\,I_{\ast}$
and (b) $I_{sc} = 0$.
 }
\end{center}
\end{figure}

\end{document}